\documentclass[journal=jctcce, layout=twocolumn, manuscript=article]{achemso}

\usepackage[T1]{fontenc}
\usepackage{amsmath} 
\usepackage{amsfonts}
\usepackage{tabularx}
  \newcolumntype{Y}{>{\centering\arraybackslash}X}

\newcommand*{\otimesg}{\ensuremath{\otimes_{\mathfrak{g}}}}
\newcommand*{\order}[1]{\ensuremath{\mathcal{O}\left(#1\right)}}

\newcommand*{\cuo}{\ensuremath{[\mathrm{Cu}_2\mathrm{O}_2]^{2+}}}

\author{Klaas Gunst}
\affiliation[Ghent University]
{Center for Molecular Modeling, Ghent University, Technologiepark 903, 9052 Zwijnaarde, Belgium}
\alsoaffiliation[Ghent University]
{Department of Physics and Astronomy, Ghent University, Krijgslaan 281, S9, B-9000 Ghent, Belgium}
\email{Klaas.Gunst@UGent.be}
\author{Frank Verstraete}
\affiliation[Ghent University]
{Department of Physics and Astronomy, Ghent University, Krijgslaan 281, S9, B-9000 Ghent, Belgium}
\alsoaffiliation[University of Vienna]
{Vienna Center for Quantum Technology, University of Vienna, Boltzmanngasse 5, 1090 Vienna, Austria}
\author{Sebastian Wouters}
\affiliation[Brantsandpatents]
{Brantsandpatents, Pauline van Pottelsberghelaan 24, 9051 Sint-Denijs Westrem (Ghent), Belgium}
\author{\"{O}rs Legeza}
\affiliation[Wigner Research Centre for Physics]
{Strongly Correlated Systems "Lend\"{u}let" Research group,
Wigner Research Centre for Physics, H-1525, Budapest, Hungary}
\author{Dimitri Van Neck}
\affiliation[Ghent University]
{Center for Molecular Modeling, Ghent University, Technologiepark 903, 9052 Zwijnaarde, Belgium}

\title{T3NS: three-legged tree tensor network states} 

\begin{document}

\begin{abstract}
  We present a new variational tree tensor network state (TTNS) ansatz, the three-legged tree tensor
  network state (T3NS). Physical tensors are interspersed with branching tensors. Physical tensors
  have one physical index and at most two virtual indices, as in the matrix product state (MPS)
  ansatz of the density matrix renormalization group (DMRG). Branching tensors have no physical
  index, but up to three virtual indices. In this way, advantages of DMRG, in particular a low
  computational cost and a simple implementation of symmetries, are combined with advantages of
  TTNS, namely incorporating more entanglement. Our code is capable of simulating quantum chemical
  Hamiltonians, and we present several proof-of-principle calculations on LiF, N$_2$ and the
  bis($\mu$-oxo) and $\mu - \eta^2 : \eta^2$ peroxo isomers of \cuo.
\end{abstract}

\section{Introduction}

Since its formulation in 1992 by S. White\cite{White1992, White1993}, the density matrix
renormalization group (DMRG) method has quickly proved its usefulness in the simulation of strongly
correlated quantum systems, both in condensed matter physics and theoretical chemistry. While
initially applied on systems with local Hamiltonians, it didn't take long before it was applied 
successfully on systems with long-range interactions, like in momentum space representation
(k-DMRG)\cite{Xiang1996} and quantum chemistry (QC-DMRG)\cite{White1999}. 

Later on, it was found that DMRG corresponded with the variational optimization of a particular wave
function, the matrix product state (MPS)\cite{Ostlund1995, Rommer1997}. In an MPS, the state is
represented by a linear chain of tensors, providing a very efficient parametrization of states
respecting the area law for entanglement in 1D systems.  This explained the high efficiency of DMRG
for the description of ground states of one-dimensional non-critical local Hamiltonians. It also
clarified the connection with quantum information theory, and paved the way to more advanced wave
functions, the so-called tensor network states (TNS).

Changing from one dimensional MPSs to other types of TNSs allowed efficient descriptions of
entanglement in higher dimensions and in critical systems. Because of this, these methods are
gaining more and more momentum, especially in condensed matter physics. Notable examples of more
general TNSs are, for example, the projected entangled pair states (PEPS)\cite{Verstraete2004} and
the multi-scale entanglement renormalization ansatz (MERA)\cite{Vidal2007}.

In quantum chemistry, other tensor network states have also been studied. Notable examples are tree
tensor network states (TTNS)\cite{Murg2010, Nakatani2013, Murg2015}, which are the subject of this
paper, complete-graph tensor network states (CGTNS)\cite{Marti2010} and self-adaptive tensor network
states (SATNS)\cite{Kovyrshin2017}. However, the MPS is still the preferred tensor network for
quantum chemistry, although its one-dimensional nature is far from ideal save for linear molecules.
Due to the high efficiency and stability of the algorithm and the relative ease of implementing
$SU(2)$-symmetry, the suboptimal entanglement representation can be lifted by increasing the virtual
bond dimension enough.

In this paper, we use the three-legged tree tensor network state (T3NS), a subclass of TTNSs. We
believe it is able to represent the entanglement of a general molecule more accurately while still
being computationally efficient. In this subclass, we also expect that the implementation of
$SU(2)$-symmetry will be no more difficult than for DMRG, which is an important prerequisite for
obtaining a highly accurate and efficient algorithm in quantum chemistry.

The paper is structured as follows. In section~\ref{sec:TTNS}, general tree tensor networks are
briefly explained and an overview is given of previous research in TTNS for quantum chemistry
(QC-TTNS). The T3NS is defined in subsection~\ref{sec:ourTTNS}, followed by a short explanation of
the fermionic sign handling in subsection~\ref{sec:fermi}. The complexity of the algorithm, some of
the most intensive steps in the algorithm and factors that influence speed and accuracy are
discussed in subsections~\ref{sec:complexity} and \ref{sec:orb}. In section~\ref{sec:results},
calculations for different quantum mechanical systems are discussed using T3NS. Summary and
conclusions are provided in section~\ref{sec:conclusion}.

This paper is meant for the reader already familiar with TNS, DMRG and more particularly QC-DMRG.
For a thorough study of these subjects we refer to refs.~\citenum{White1992, White1993, White1999,
  Chan2002, Legeza2003b, Schollwock2011, Orus2014, Szalay2015, Mcculloch2002, Sharma2012,
  Wouters2014a, Wouters2014b, Keller2015, Chan2016}.

\section{\label{sec:TTNS}Tree Tensor Networks}

The TTNS is a natural extension of the MPS ansatz which is used in DMRG. While the MPS wave function
can be depicted as a linear chain of tensors, the TTNS ansatz allows branching of the network. The
TTNS ansatz is the most general tensor network state without any loops. It allows an exact treatment
from the mathematical point of view as higher order singular value decomposition (HOSVD) can be
applied. By using this ansatz, a better representation of the entanglement topology of the system is
expected as compared to the MPS, since component tensors can have an arbitrary order.

A substantial advantage of TTNS is that at a finite bond dimension it is able to capture
algebraically decaying correlation functions. This in contrast to DMRG which is only able to
represent exponentially decaying correlations\cite{Murg2010,Murg2015,Wouters2014b}. This can easily
be seen as follows. Imagine we start from one central tensor and we radially expand the TTNS with a
fixed coordination number $z$ (i.e. the maximum number of virtual bonds of a tensor in the TTNS).
The number of sites $L$ in function of the number of layers $Y$ is
\begin{equation}\label{eq:treesize}
  L = 1 + z \sum_{k=1}^Y (z - 1)^{k - 1} = \frac{z(z-1)^Y - 2}{z-2}
\end{equation}
for $z \geq 3$ or
\begin{equation}\label{eq:mpssize}
  L = 1 + 2 Y 
\end{equation}
for $z = 2$ which is the MPS case. The maximal distance between two sites is given by $2Y$. From
eq.~(\ref{eq:treesize}) and eq.~(\ref{eq:mpssize}) follows a logarithmic scaling of maximal distance
with system size $L$ for trees and a linear one for MPSs. Correlation functions in TTNSs with finite
bond dimension decay exponentially in function of maximal distance. Hence, in function of system
size an algebraic decay is obtained for $z \geq 3$ in contrast to the exponential decay for the MPS
$(z=2)$\cite{Shi2006,Ferris2013,Wouters2014b}.

Tree tensor networks for quantum chemistry (QC-TTNS) were first studied by Murg \latin{et
al.}\cite{Murg2010,Murg2015} for trees with arbitrary coordination number. The complexity of the
algorithm as a function of the virtual dimension $D$ is given by $\order{D^{x+1}}$, where $x$ is
given by the coordination number of the tensor optimized at each stage. Due to this scaling, Murg
\latin{et al.} restricted themselves to a maximum coordination number of 3 in the network and to a
one-site optimization scheme which results in $\order{D^4}$.  A two-site optimization scheme in a
tree with coordination number of 3 includes optimizing two-site tensors with 4 virtual bonds as can
be seen in fig.~\ref{fig:TTNS}(a). This ultimately results in an expensive $\order{D^5}$.

In DMRG, the usage of a two-site optimization scheme has proved to be advantageous. The two-site
scheme is less prone to be stuck in local minima and an automatic redistribution of the virtual
dimensions over different symmetry sectors is possible through singular-value decomposition
(SVD)\cite{Legeza2003b}. In TTNS it would be opportune to also use two-site optimization. In contrast
to DMRG though, two-site optimization for an arbitrary TTNS is accompanied with a heavier polynomial
cost than one-site optimization as previously stated. In the work of Nakatani \latin{et
al.}\cite{Nakatani2013} this problem is circumvented by introducing \textit{half-renormalization}.
In the half-renormalization step the TTNS is exactly mapped to an MPS. In this MPS, the iterative
optimization step is executed at a DMRG-like cost. The mapping of the TTNS to an MPS is still
expensive, though.

In this paper, we propose the T3NS ansatz which has considerable advantages compared to a general
TTNS. We show that the proposed ansatz enables two-site optimization without any penalty in the
polynomial scaling and without the need of mapping through the half-renormalization scheme.

\subsection{\label{sec:ourTTNS} The T3NS ansatz}

\begin{figure}[h!]
  \centering
  \includegraphics[width=\columnwidth]{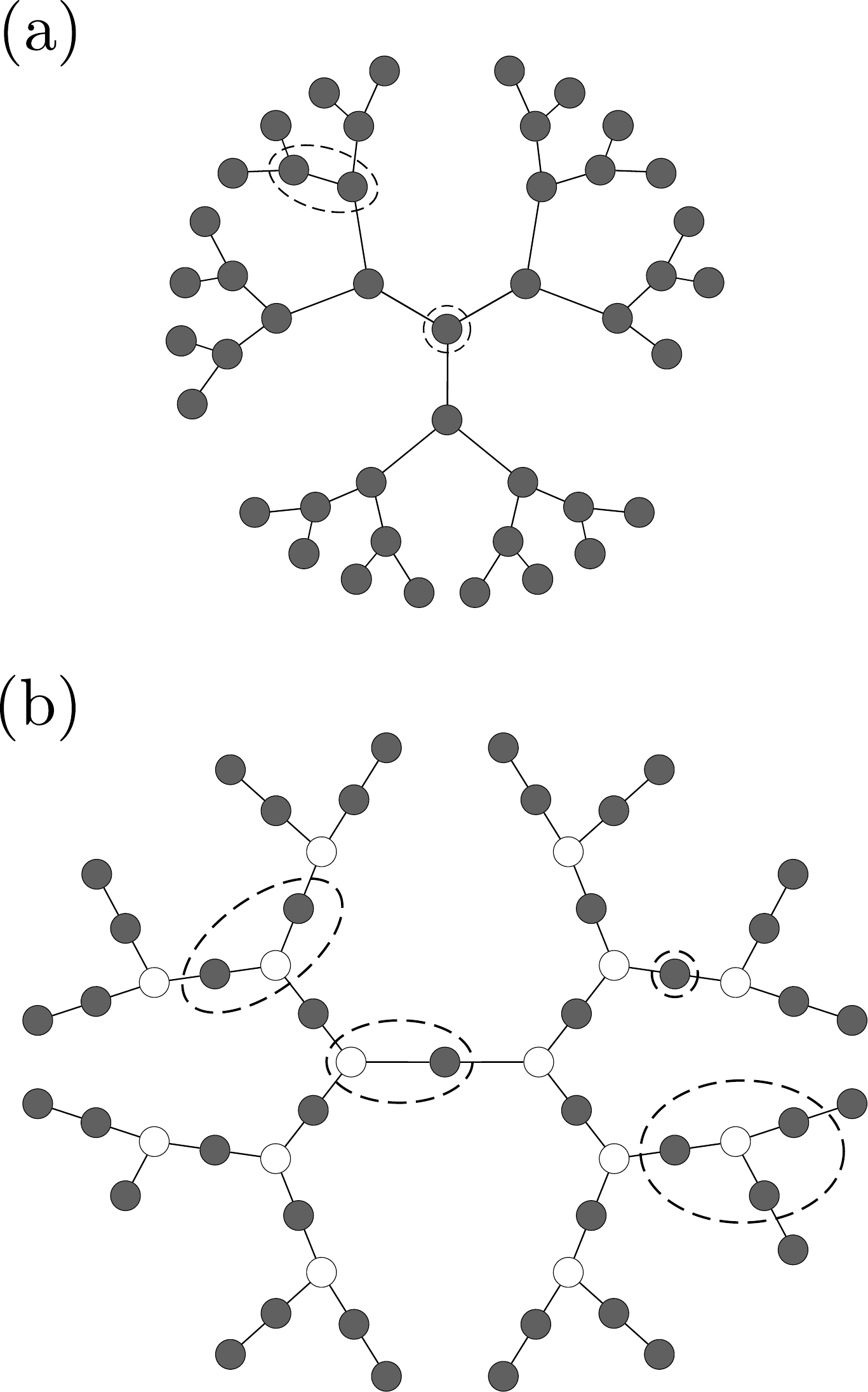}
  \caption{\label{fig:TTNS} (a) A general TTNS with maximum $z=3$ and 44 orbitals. The tensor
    optimized during one or two-site optimization is shown by dashed lines, and has maximally 3 or 4
    virtual bond indices, respectively. (b) An example of a T3NS with 44 orbitals.  The tensor
    optimized during one, two-, three-, or four-site optimization is shown by the dashed contours,
    and has maximally 3 virtual bond indices for all cases. Filled circles represent
    \textit{physical tensors} and have thus an extra physical index that is not drawn here for
    simplicity. Empty circles represent \textit{branching tensors}.}
\end{figure}

Just as in previous works on QC-TTNS\cite{Murg2010, Nakatani2013, Murg2015}, we restrict ourselves
to a maximum coordination number $z=3$, to keep calculations feasible. A second restriction we
impose, is that only tensors with $z \leq 2$ have physical indices. We call this type of tensors
\textit{physical tensors}. Tensors with $z=3$ are called \textit{branching tensors} and have
exclusively virtual bond indices. An example of this type of TTNS (the T3NS) is given in
fig.~\ref{fig:TTNS}(b).

The proposed ansatz enables us to go beyond one-site optimization and use two-site, three-site or
even four-site optimization with the same polynomial scaling (see fig.~\ref{fig:TTNS}(b)). Another
substantial advantage of T3NS is that every tensor has at most three different indices (one physical
and two virtual for a physical tensor and three virtual for a branching tensor). Hence no extra
substantial difficulties are expected for the implementation of $SU(2)$-symmetry, compared with the
MPS formalism\cite{Mcculloch2002, Toth2008, Sharma2012, Wouters2014a, Wouters2014b}.

At this moment our T3NS implementation is able to do two-site, three-site and four-site optimization
and it exploits $U(1)$-symmetry. In this paper, only two-site optimization is used. Three- and
four-site calculations were executed but the small increase in energy accuracy did not outweigh the
extra computational time needed (larger prefactor). However, the ability to do three- and four-site
optimization can be useful for orbital optimizations. The usage of $SU(2)$-symmetry will be the
subject of subsequent research.

\subsection{\label{sec:fermi}Fermionic Networks}

A quantum chemical calculation involves fermions. This introduces extra complexity in the algorithm
through the sign change of the wave function when interchanging two fermions. For our implementation
we opted for the fermionic network formalism as developed by Bultinck \latin{et al.}\cite{Bultinck2017}

In this formalism, fermionic tensors are given by
\begin{align}
  A = \sum_{\alpha \beta \gamma \delta \dots} A_{\alpha \beta \gamma \delta \dots} 
  |\alpha) |\beta) (\gamma| (\delta| \dots 
\end{align}
This is equivalent with the definition for bosonic tensors, but where $|\alpha),|\beta),(\gamma|$,
$(\delta|,\dots$ are instead elements of the so-called super vector space $V$.  Bras can be
graphically depicted as outgoing tensor legs, while kets are ingoing tensor legs.

The fermionic signs are introduced by the following canonical isomorphism
\begin{align}\label{eq:pmap}
  \mathcal{F}:& V\otimesg W \rightarrow W\otimesg V \nonumber \\
  & |i\rangle\otimesg|j\rangle \rightarrow (-1)^{|i||j|}|j\rangle\otimesg|i\rangle,
\end{align}
where $V$ and $W$ are super vector spaces and $\otimesg$ denotes the graded tensor product.
$|i\rangle$ and $|j\rangle$ represent homogeneous basis states. A homogeneous state is characterized
by a definite parity of the state (namely, $|i|, |j|  \in \{0,1\}$). 

For the contraction of fermionic tensors, a second mapping is introduced:
\begin{equation}\label{eq:cmap}
  \mathcal{C} : V^* \otimesg V \rightarrow \mathbb{C} : 
  \langle\psi|\otimesg |\phi\rangle \rightarrow \langle\psi|\phi\rangle,
\end{equation}
where $V$ and $V^*$ are the super vector space and its dual space, respectively.

When contracting two fermionic tensors, the states of the tensors should first be ordered
appropriately through successive usage of eq.~(\ref{eq:pmap}) before using eq.~(\ref{eq:cmap}). No
explicit ordering of the orbitals is needed this way, but this is implicitly fixed by the initial
order of the indices in the different tensors of the network. For further details we refer to
ref.~\citenum{Bultinck2017}.

\subsection{\label{sec:complexity}Resource requirements of the algorithm}

In this section the computational complexity and the memory requirements of the algorithm are
discussed. For the implementation of the algorithm, we opted for the usage of (complementary)
renormalized operators, just as in previous works on QC-TTNS\cite{Murg2010, Nakatani2013, Murg2015}.
This approach for the efficient calculation of expectation values has also been heavily used in
highly optimized QC-DMRG.

Another technique, the so-called Matrix Product Operator (MPO) formalism, has also been formulated
for the quantum chemistry Hamiltonian. In this formalism, the Hamiltonian is represented by a tensor
network too, consisting of different MPOs (as opposed to MPSs for the wave function). To obtain an
efficient MPO representation of the Hamiltonian, the bond dimension of the MPO should be at least of
the same order as the number of renormalized operators used, i.e.  $\order{k^2}$, with $k$ the
number of orbitals. When such a representation is found, one still has to exploit the extra sparsity
of the MPOs to obtain the same cost as with renormalized operators.  For QC-DMRG, several methods
of obtaining such MPO and exploiting the extra sparsity are already known\cite{Keller2015,Chan2016}.

Equivalently for QC-TTNS, the QC-Hamiltonian can be formulated in the Tensor Network Operator (TNO)
language. While the Hamiltonian is represented by a linear tensor network in the MPO formalism, the
Hamiltonian can also be represented by a tree tensor network through usage of TNO's. Again, a first
condition for an efficient TNO is a scaling of $\order{k^2}$ for the bond dimension. Just as in DMRG
exploitation of sparsity is needed to obtain similar costs as with renormalized operators. Failing
to do so is even more catastrophic for QC-TTNS than for QC-DMRG. The methods proposed in
refs.~\citenum{Keller2015, Chan2016} are not readily translatable to QC-TTNS or do not produce the
same scaling as with renormalized operators. Because of this, we opted for the usage of renormalized
operators.

\begin{table}[h!] 
  \centering 
  \resizebox{\columnwidth}{!}{%
    \begin{tabular}{r c | c} & DMRG & T3NS \\
    \cline{2-3} & &\\
    CPU time: & $\order{k^4D^2 + \underline{k^3D^3}}$  & $\order{k^5D^2 + \underline{k^3D^4}}$\\
    Memory:   & $\order{k^2D^2}$                       & $\order{k^2D^2 + kD^3}$\\
    Disk:     & $\order{k^3D^2}$                       & $\order{k^3D^2 + kD^3}$
  \end{tabular}}

  \caption{\label{tab:req}Resource requirements of DMRG and T3NS with renormalized operators.  The
    underlined terms correspond with the complexity of the most intensive part of the algorithm,
    i.e. the matrix-vector product used in the iterative solver.} 
\end{table}

The predicted scaling of CPU time, memory usage, and disk usage are given in Table~\ref{tab:req} and
compared with DMRG. The most time consuming part of the algorithm is the iteratively executed
matrix-vector product of the effective Hamiltonian with the two-site tensor ($H_{\mathrm{eff}}
\Psi$).  Due to the usage of complementary renormalized operators, the effective Hamiltonian is
constructed out of $\order{k^2}$ different terms, for both DMRG and T3NS.  However, the cost of
constructing each term scales as $\order{D^4}$ for T3NS instead of $\order{D^3}$ for DMRG. 

\begin{figure}[h!] 
  \centering 
  \includegraphics[width=\columnwidth]{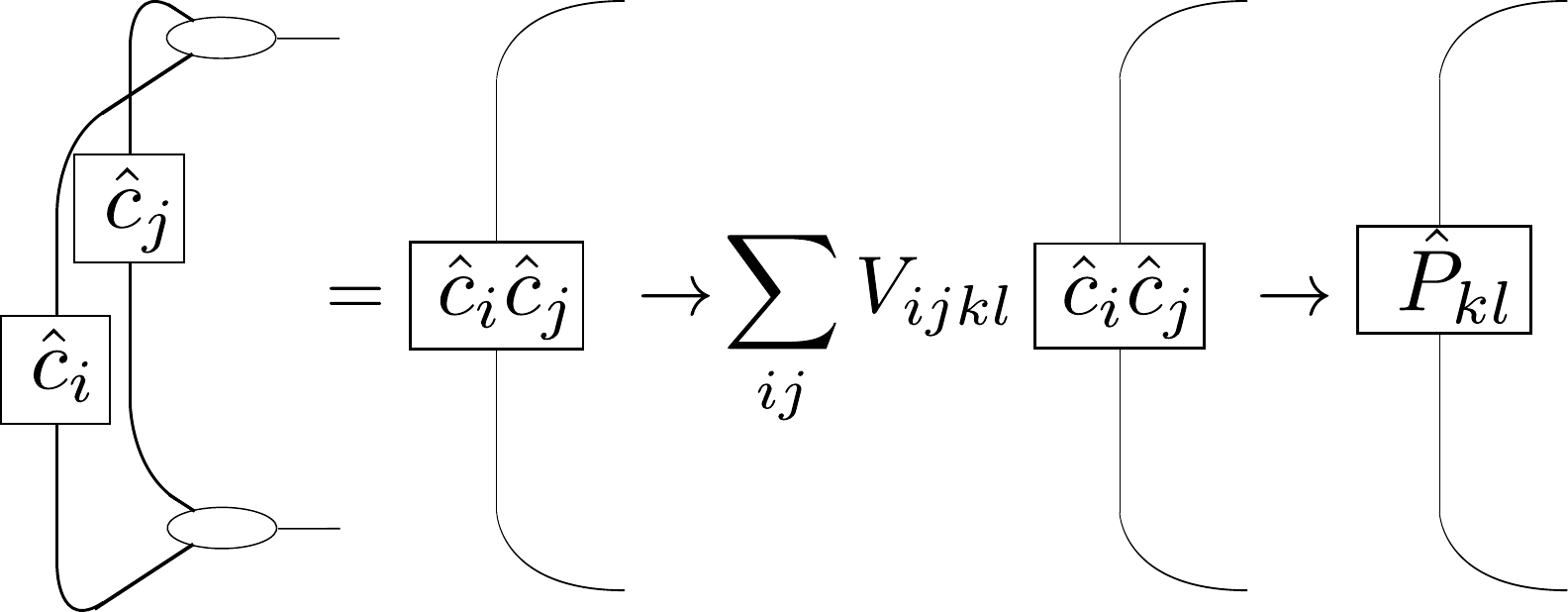}
  \caption{\label{fig:update} A graphical depiction of the most intensive part of updating the
    renormalized operators, i.e. the recombination of two single operators into a double
    complementary operator. First, the single operators are combined to a double operator with the aid
    of a branching tensor $\left(\order{k^2D^4}\right)$. In the second stage, the newly formed double
    operators are summed together with their potential terms into the different double complementary
    operators $\left(\order{k^4D^2}\right)$. Since there are $\order{k}$ occasions per sweep for
    this, we obtain $\order{k^5D^2 + k^3D^4}$.} 
\end{figure}

The other leading term in the CPU time per sweep is due to updating the renormalized operators.  The
most intensive type of update for renormalized operators is when two sets of renormalized operators
have to be recombined in a new set, this by using a branching tensor. The most intensive type of
recombination occurs when a single operator in both sets has to be updated to a complimentary double
operator. This results in $\order{k^5D^2 + k^3D^4}$ per sweep, as can be seen in
fig.~\ref{fig:update}.

At a fixed system size $k$ and bond dimension $D$, the speed of a sweep is still dependent of the
particular shape of the tree. While one can only make one shape of MPS-chain for a fixed system
size, this is not true for trees. 

\subsection{\label{sec:orb}Shape of the tree, orbital ordering and choice}

As previously stated, the different shapes for the tree at fixed $k$ introduce additional freedom
that is not present in DMRG. It is clear that the particular shape will influence the speed and the
accuracy of the calculations. The orbital ordering in the network is also of importance for the
accuracy of the calculations. This freedom is also present in DMRG and multiple methods for
ordering the orbitals exist (e.g. through use of the mutual information\cite{Murg2015} or the
exchange integral\cite{Nakatani2013} have been studied). Similar methods can be used to optimize
the shape of the tree.

Finally, the orbital choice and orbital optimization is also of importance for TTNS and DMRG
calculations. Quite some research has been done for this in DMRG\cite{Legeza2003, Rissler2006,
  Ghosh2008, Zgid2008, Barcza2011, Krumnow2016}. In TTNS, orbital optimization by canonical
transformations has been studied and used\cite{Murg2010}.

In this paper, we group orbitals belonging to the same spatial irrep as much as possible and connect
the irreps in the center of the tree. The used trees and orbital orderings are given in the
supplementary material. Within one irrep, the orbitals are ordered such that the orbitals closest
to the Fermi level (for LiF and N$_2$) or the orbitals with highest single-orbital entropy (for
\cuo) are closest to the center of the tree. After experimenting with a few different orderings of
the orbitals, this proved to be the most successful one. Another degree of freedom is the choice of
the orbitals. In this paper, we only use Hartree-Fock orbitals. No orbital optimization is
executed, as this paper serves only as an initial description of our particular T3NS ansatz.

Optimization of the shape of the network, the orbital order, or the orbitals themselves will be the
subject of subsequent research.

\section{\label{sec:results}Numerical results}

In this section, we compare the T3NS ansatz with the MPS ansatz. Energy errors and CPU times are
compared in function of the bond dimension. We study LiF and $\mathrm{N}_2$ at their equilibrium
bond length ($r = 3.05$ a.u. and $r = 2.118$ a.u., respectively). For LiF we also calculations at $r
= 12$ a.u. and $r=13.7$ a.u. LiF and N$_2$ are two systems that don't particularly call for a
tree-shaped topology representation. However, as we will show, a similar accuracy is already
obtained with the T3NS at considerable lower bond dimension as compared with DMRG for both systems.
This fact gives us hope that the more complex entanglement topology will prove even more its merits
in larger molecules, since the orbitals can be easier arranged in groups of highly entangled
orbitals\cite{Szalay2015}.

LiF and $\mathrm{N}_2$ are studied for different bond dimensions by T3NS with $U(1)$-symmetry and
DMRG with $U(1)$-symmetry. LiF is also studied with DMRG with $SU(2)$-symmetry. For both T3NS-$U(1)$
and DMRG-$U(1)$ we use our own implementation of the T3NS ansatz. It is of course also possible to
do DMRG-$U(1)$ since this is just a subset of the possible tree-shaped geometries. For DMRG-$SU(2)$
we use the CheMPS2 software program developed by S. Wouters\cite{Wouters2014a, Wouters2014b,
  Wouters2014c, Wouters2016}.

Both systems are popular benchmarks for methods and their ability to take strong electron
correlations into account, and both systems have been studied in previous papers about QC-TTNS (LiF
in ref.~\citenum{Murg2015} and $\mathrm{N}_2$ in ref.~\citenum{Nakatani2013}).

To test the T3NS ansatz in larger systems, we perform calculations on the bis($\mu$-oxo) and the
$\mu - \eta^2 : \eta^2$ peroxo \cuo isomers. We compare the energy gaps between the two isomers
obtained by T3NS with previously published values\cite{Cramer2006, Malmqvist2008, Marti2008,
  Kurashige2009, Yanai2010, Marti2010b, Barcza2011, Stein2016, Phung2016}. We also compare our
complete active space (CAS) ground-state energies with other calculations executed in the same
active space. This way we keep comparisons fair.

\subsection{Results for LiF}

\begin{figure}[h!] 
  \centering
  \includegraphics[width=\columnwidth]{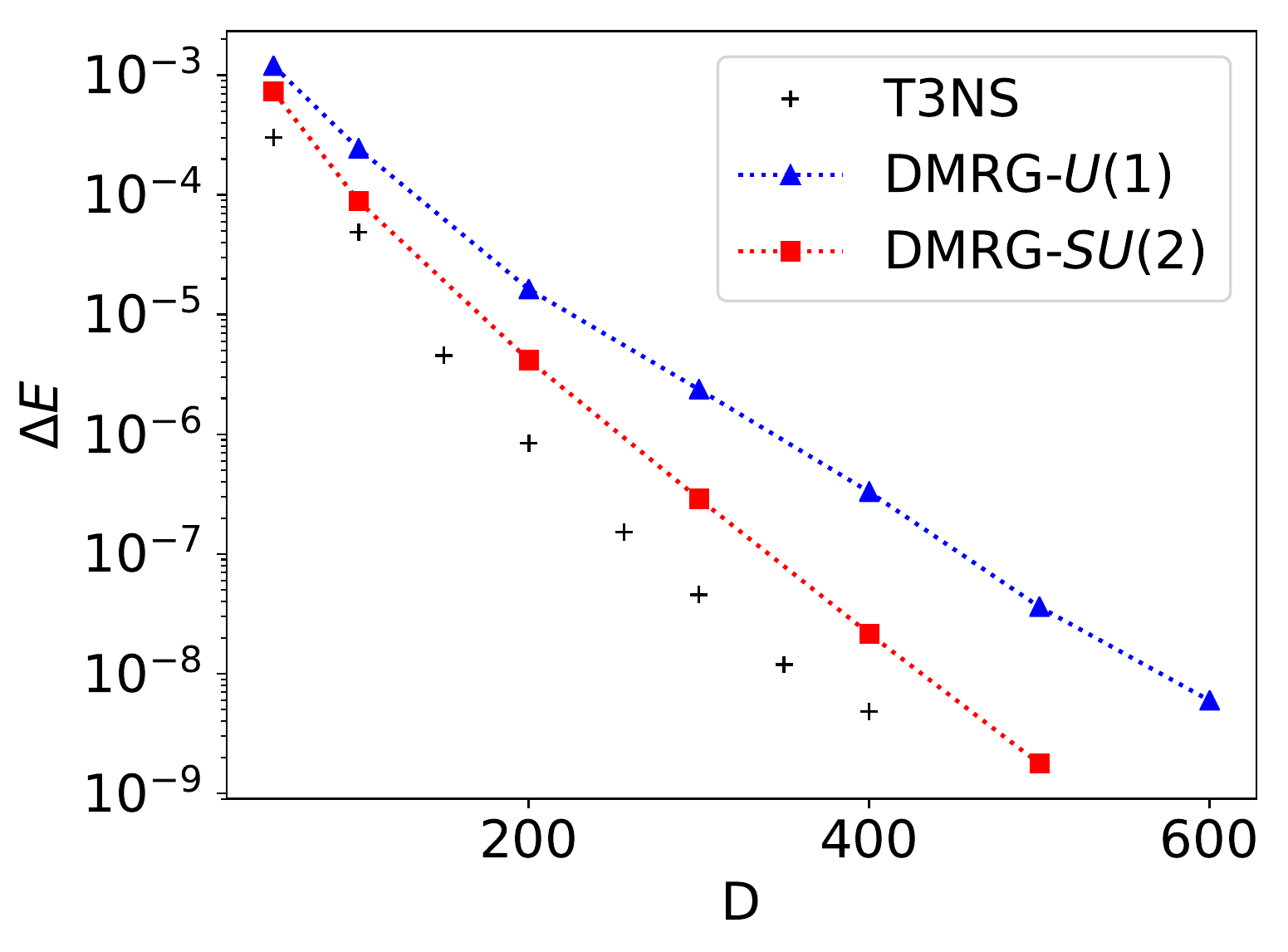} 
  \caption{\label{fig:LiFe} Energy difference of DMRG and T3NS calculations with respect to the FCI
    energy for LiF at equilibrium bond length $r = 3.05$ a.u. FCI energies are obtained
    from ref.~\citenum{Murg2015}. The calculations are done at different (\textit{reduced}) bond
    dimensions. The TTNS geometry is given in the supplementary material.}
\end{figure}

\begin{figure}[h!] 
  \centering
  \includegraphics[width=\columnwidth]{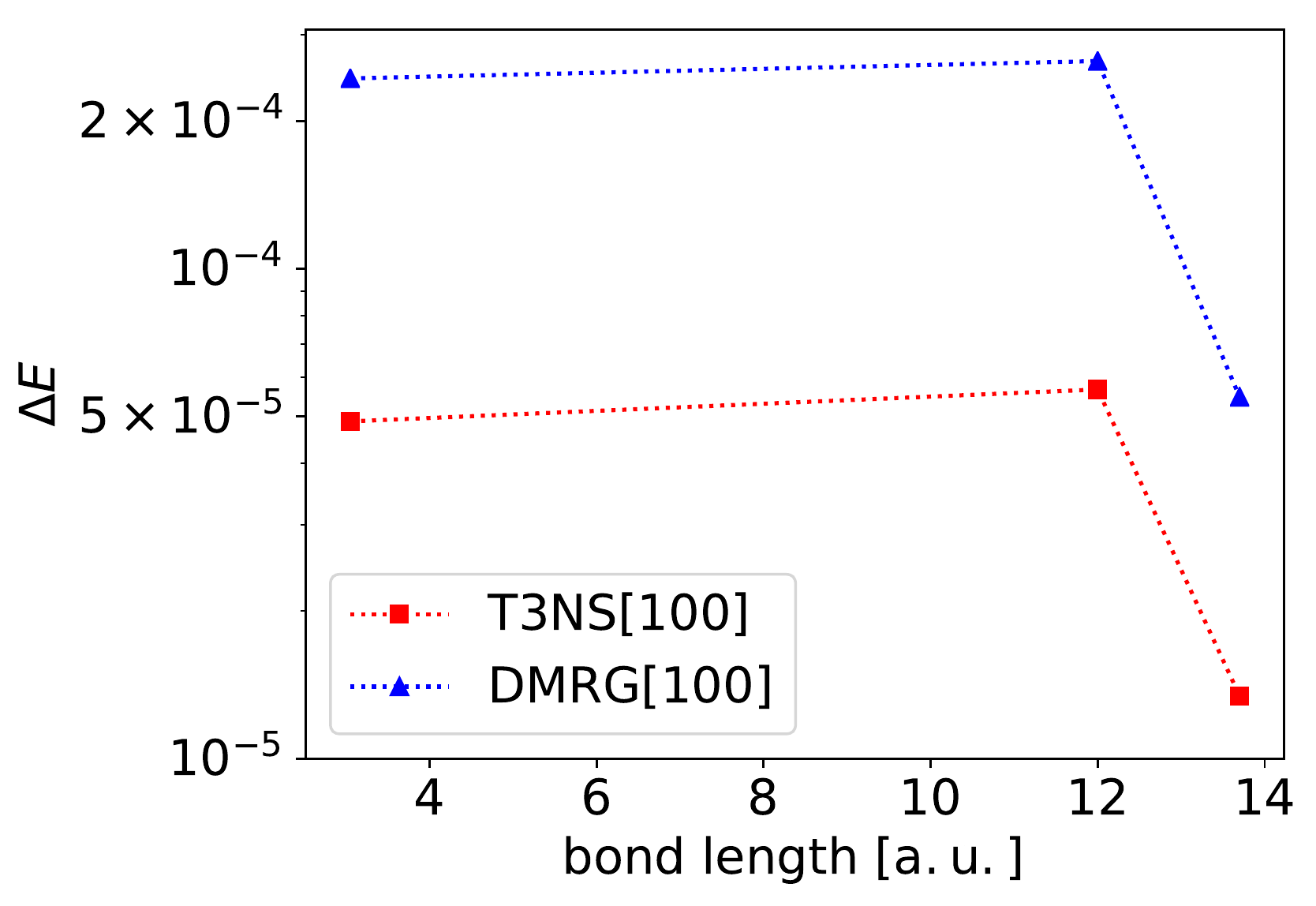}
  \caption{\label{fig:LiFbl} Energy difference of DMRG-$U(1)$ and T3NS-$U(1)$ calculations with
    respect to the FCI energy for LiF at bond length $r =$ 3.05, 12 and 13.7 a.u. FCI energies are
    obtained from ref.~\citenum{Murg2015}. The calculations are done at $D=100$ for both DMRG and T3NS.
    The TTNS geometry is given in the supplementary material.}
\end{figure}

\begin{table}[h!] 
  \centering 
  \resizebox{\columnwidth}{!}{%
    \begin{tabular}{l |  c | c} & CPU time last sweep & total CPU time \\
      \hline
    T3NS[100] & 96 sec & 428 sec \\
    T3NS[400] & 1000 sec & 2240 sec \\
         \hline
    DMRG[100] & 48 sec  & 600 sec \\
    DMRG[600] & 640 sec & 1884 sec \\
  \end{tabular}}

  \caption{\label{tab:timeLiF}Some timings for T3NS and DMRG calculations of LiF at equilibrium bond
    length. Used bond dimensions are given in square brackets. Both T3NS and DMRG are executed
    with our own implementation to keep comparison fair.}
\end{table}

The first system we study with the new T3NS ansatz is LiF. We perform calculations at equilibrium
bond length $(r = 3.05$ a.u.), at $r = 12$ a.u. where an avoided crossing occurs, and at large bond
length $(r =13.7$ a.u.). The bond lengths are expressed in atomic units. Calculations are performed
in a CAS of size (6e, 25). The atomic orbital basis from Bauschlicher and
Langhoff\cite{Bauschlicher1988} was used. For the active space calculations, the $1\sigma$,
$2\sigma$ and $3\sigma$ orbitals were kept frozen. The same basis set and active space is used in
ref.~\citenum{Murg2015}. Ground state energies were calculated by using T3NS-$U(1)$, DMRG-$U(1)$ and
DMRG-$SU(2)$  with several bond dimensions. In the case of DMRG-$SU(2)$, the quoted bond dimension
is the reduced one, where the additional $SU(2)$-symmetry is taken into account.\cite{Wouters2014a}
FCI energies were easily recovered through the T3NS. Accuracies in the order of $10^{-8}$
$\mathrm{E_h}$ were obtained for all bond lengths at $D=400$.  The accuracy of DMRG and T3NS in
relation to the bond dimension is given in fig.~\ref{fig:LiFe} for LiF in equilibrium. As expected,
a lower bond dimension is needed for T3NS for a similar accuracy as in DMRG. In
fig.~\ref{fig:LiFbl}, the accuracy of DMRG and T3NS is given for different bond lengths at a low
bond dimension $(D=100)$.

Lastly, some wall times for the T3NS and DMRG calculations are given in table~\ref{tab:timeLiF} for
$r = 3.05$.  At $D=100$, a sweep is twice as slow in T3NS as in DMRG, as can be expected. However,
less sweeps are needed until convergence which ultimately results in a faster calculation with
higher accuracy. The need for fewer sweeps in T3NS is something we noticed quite consistently. For
DMRG at $D=600$ and T3NS at $D=400$ both accuracy and total wall time are comparable. We would like
to note that these remarks on timing are by no means conclusive since the speed and accuracy of both
T3NS and DMRG are heavily dependent on orbital ordering and initial guess. In these calculations, a
random initial guess and a rather intuitive orbital ordering was used. These remarks are merely to
illustrate the competitiveness of our T3NS ansatz with DMRG.

The used TTNS geometries and orbital orderings are given in the supplementary material.

\subsection{Results for $\mathrm{\textbf{N}}_\mathbf{2}$}

\begin{figure}[h!] 
  \centering
  \includegraphics[width=\columnwidth]{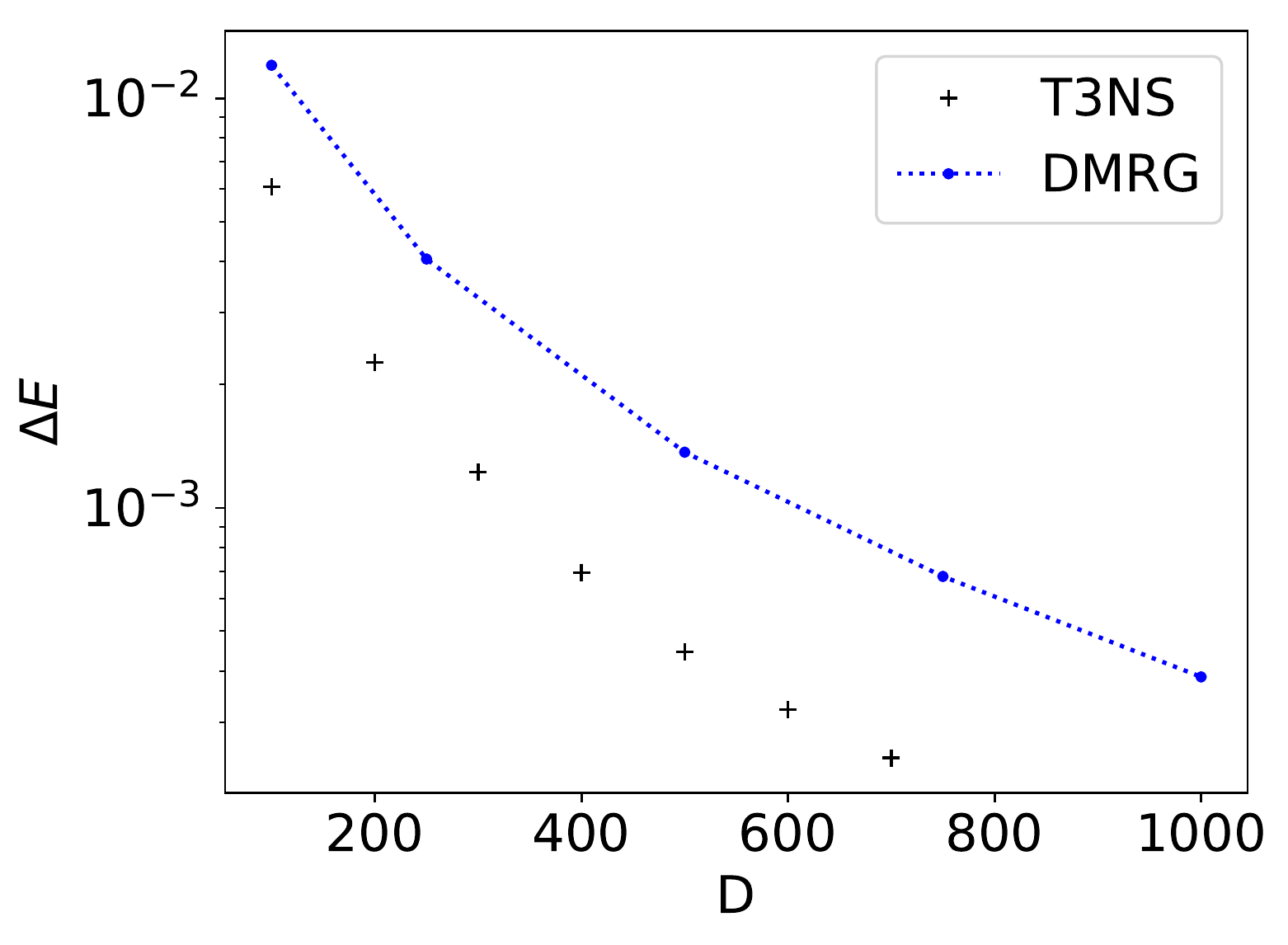}
  \caption{\label{fig:N2e} Energy difference of DMRG and T3NS calculations for $\mathrm{N}_2$ with
    respect to CCSDTQPH energy (-109.282172 $\mathrm{E_h}$)\cite{Chan2004}. The calculations are
    done at different bond dimensions. The TTNS geometry is given in the supplementary material.}
\end{figure}

\begin{table}[h!] 
  \centering 
  \resizebox{\columnwidth}{!}{%
    \begin{tabular}{l |  c | c} & CPU time last sweep & total CPU time \\
         \hline
    T3NS[100] & 560 sec & 1440 sec \\
    T3NS[300] & 4h & 17h \\
    T3NS[500] & 16h & 96h \\
    T3NS[700] & 66h & 237h \\
         \hline
    DMRG[100] & 160 sec  & 3800 sec \\
    DMRG[500] & 2050 sec & 9h \\
    DMRG[1000] & 2h & 27h \\
  \end{tabular}}

  \caption{\label{tab:timeN2}Some CPU times for T3NS and DMRG calculations of N$_2$ at equilibrium
    bond length. Both T3NS and DMRG are executed with our own implementation to keep comparison
    fair. Ordering of the orbitals on the network are given in the supplementary material.} 
\end{table}

The second benchmark system for our T3NS ansatz is the nitrogen dimer at equilibrium (bond
length: 2.118 a.u.). This is a popular molecule for benchmarking methods in their ability to
describe strong electron correlation accurately. Because of this it has also been discussed by
Nakatani \latin{et al.} in their TTNS paper\cite{Nakatani2013}. They studied the nitrogen dimer in a
cc-pVDZ basis set in a frozen core active space (10e, 26), keeping the 1s electrons of nitrogen
fixed.  DMRG\cite{Chan2004} and FCI\cite{Larsen2000} calculations have also been previously executed
for this active space. In this paper we execute all-electron calculations (14e, 28) for the nitrogen
dimer in a cc-pVDZ basis set and compare them with the most accurate results obtained in
ref.~\citenum{Chan2004} through coupled cluster on the SDTQPH level.

Several calculations have been executed at different bond dimensions for T3NS and DMRG with $U(1)
\times U(1)$-symmetry. The obtained energy differences with respect to CCSDTQPH\cite{Chan2004} are
given in fig.~\ref{fig:N2e} for bond dimensions up to 1000 for DMRG and up to 700 for T3NS.
Comparable energies are obtained for T3NS at half the bond dimension needed for DMRG. This is
consistent with the conclusion from the frozen core TTNS calculations in ref.~\citenum{Nakatani2013}.

CPU times are given in table~\ref{tab:timeN2} for T3NS-$U(1)$ and DMRG-$U(1)$ calculations. Similar
conclusions can be made in comparison with LiF. For $D=100$, T3NS-sweeps take longer than
DMRG-sweeps, but the number of sweeps needed for convergence from a random initial guess is
considerably lower. This ultimately results in a lower wall time. At $D=1000$  for DMRG and $D=500$
for T3NS, obtained accuracies are comparable. Wall times for T3NS are higher though than for DMRG,
but still in the same order.

\subsection{Results for the bisoxo and peroxo isomer of 
  $\mathbf{[\mathrm{\textbf{Cu}}_2\mathrm{\textbf{O}}_2]^{2+}}$}

As a last benchmark system, we study the bisoxo($\mu$-oxo) and peroxo isomers of \cuo, and in
particular their energy gap.  These transition metal clusters have been studied with a wide range of
ab initio methods like CASSCF and CASPT2 (complete active space self consistent field theory with
perturbation theory up to second order)\cite{Cramer2006} and RASPT2 (restricted active space self
consistent field theory with perturbation theory up to second order)\cite{Malmqvist2008}.  However,
the small active spaces used for CASPT2 and RASPT2 showed to be insufficient. Later on, the usage of
DMRG-based methods allowed to take a considerably larger active space into account, yielding
improved results\cite{Marti2008, Kurashige2009, Yanai2010, Marti2010b, Barcza2011, Stein2016,
  Phung2016}.

In this paper, we use the T3NS algorithm to treat the two isomers in a (26e, 44) active space. We
use the same active space and basis set as in ref.~\citenum{Marti2008} and \citenum{Barcza2011}.
Results are given in table~\ref{tab:cu2o2}. Energies of the isomers in the used active spaces are
very comparable to the ones in ref.~\citenum{Barcza2011}. Furthermore, the energy gap between the
two isomers are in the same region as previously executed DMRG calculations.

\begin{table*}[t] 
  \centering 
  \begin{tabularx}{\textwidth}{ l Y Y r }
    \hline\hline
    Ref. Method & $E_\mathrm{bisoxo}[\mathrm{E_h}]$ & $E_\mathrm{peroxo}[\mathrm{E_h}]$ & $\Delta E$
    [kcal/mol]\\
    \hline
    \citenum{Cramer2006} CASSCF(16,14) & & & 0.2\\
    \citenum{Cramer2006} CASPT2(16,14) & & & 1.4\\
    \citenum{Malmqvist2008} RASPT2(24,28) & & & 28.7\\
    \\
    \multicolumn{4}{c}{Some previously published DMRG energies} \\
    \citenum{Kurashige2009} DMRG(32,62)[2400] & & & 35.6\\
    \citenum{Yanai2010} DMRG(28,32)[2048]-SCF/CT & & & 27.0\\
    \citenum{Stein2016} DMRG(32,28)[4000] & & & 21.8\\
    \citenum{Phung2016} DMRG(24,24)[1500]-SCF$^*$ & & & 35.1\\
    \citenum{Phung2016} DMRG(24,24)[1500]-CASPT2$^*$ & & & 23.2\\
    \citenum{Marti2008} DMRG(26,44)[800] & -541.46779 & -541.49731 & 18.5\\
    \citenum{Marti2010b} DMRG(26,44)[128] & -541.47308 & -541.51470 & 26.1\\
    \citenum{Barcza2011} DMRG(26,44)[256/1024/$10^{-5}$]$^\dagger$ & -541.53853 & -541.58114 & 26.7\\
    \\
    \multicolumn{4}{c}{T3NS calculations} \\
    T3NS(26,44)[50] & -541.48773 & -541.56999 & 51.6\\
    T3NS(26,44)[100] & -541.52352 & -541.57166 & 30.2\\
    T3NS(26,44)[200] & -541.53284 & -541.57717 & 27.8\\
    T3NS(26,44)[300] & -541.53556 & -541.57966 & 27.7\\
    T3NS(26,44)[500] & -541.53820 & -541.58094 & 26.8\\
    \hline\hline

    \multicolumn{4}{l}{$*$: Bond dimensions given for these calculations are \textit{reduced} bond
      dimensions.}\\
    \multicolumn{4}{l}{\parbox{0.97\textwidth}{$\dagger$: This calculation uses the DBSS method and
        CI-DEAS as initialization procedure.  The square brackets state that a minimum of $D=256$ is
        used at every bond, the CI-DEAS procedure starts with $D=1024$ and a maximum discarded
        weight of $10^{-5}$ is aimed for.  Maximum bond dimensions around 2000 were reported for
        both clusters during these calculations\cite{Barcza2011}.}}
  \end{tabularx}
  \caption{\label{tab:cu2o2} Energy gaps between the bis($\mu$-oxo) and $\mu - \eta^2:\eta^2$ peroxo
    \cuo isomers from T3NS calculations of this paper and previous calculations. The energy gaps are
    given in kcal/mol. Ground state energies are given for the T3NS calculations and DMRG
    calculations from previous research using the same active space and are given in Hartree. Bond
    dimensions used for the T3NS and DMRG calculations are given in square brackets.}
\end{table*}

These results are especially promising since no advanced methods were used to augment the T3NS
calculations in contrast to the previous research with DMRG for this system.  At this moment
our algorithm starts from a random initial guess, no effort was made in avoiding local minima and an
intuitive orbital ordering was used. In contrast, previous DMRG research included the configuration
interaction based dynamically extended active space (CI-DEAS) procedure\cite{Barcza2011} or adding
perturbative noise to the tensors\cite{Phung2016} to avoid local minima. Orbitals were ordered by
minimizing quantum entanglement using the Fiedler vector\cite{Barcza2011, Phung2016}, or a genetic
algorithm\cite{Phung2016}. Other methods used to augment the results were DMRG-SCF (self consistent
field)\cite{Yanai2010}, DMRG-SCF with canonical transformation theory (DMRG-SCF/CT)\cite{Yanai2010}
or DMRG-CASPT2 to take dynamical correlation into account. In ref.~\citenum{Barcza2011} dynamic
block state selection (DBSS) was used to tune the bond dimension. Instead of a fixed bond
dimension, a maximum discarded weight is used. In this way, the bond dimension at every bond is
tailored to stay below this maximum discarded weight. DBSS is easily implementable once two-site
optimization is used, like in our T3NS algorithm.  Since we noted that the discarded weight was
dominant in very few bonds while it was orders lower in other bonds, we think that DBSS can also
yield a substantial improvement in the T3NS algorithm.

To check if we got stuck in local minima, the ground state wave function of the bisoxo isomer
obtained through T3NS[500] was compressed to a lower bond dimension. The compressed wave function
was then reoptimized at this lower bond dimension and we found a ground state energy of -541.50527,
-541.52387 and -541.53327 Hartree for $D=$ 50, 100 and 200, respectively.  Comparison with the
results obtained through random initialization in table~\ref{tab:cu2o2} makes the problem of local
minima quite clear and shows us that preventing local minima can improve our results significantly,
especially at low bond dimension.

\section{\label{sec:conclusion}Conclusion}

In this paper, we have presented a new variational ansatz, the T3NS ansatz. This is a subclass of
the general TTNS ansatz which has considerable advantages. By interspersing physical and
branching tensors in the network, two-site optimization (and even three- and four-site optimization)
becomes feasible at the same polynomial cost as one-site optimization. Furthermore, both physical
and branching tensors in the T3NS network have at most 3 indices which allows a simple
implementation of symmetries. In this way, we join the computational efficiency of the MPS with the
richer entanglement description of the TTNS.

As a proof-of-concept, calculations were executed for LiF, N$_2$ and \cuo with our T3NS
implementation. Accuracies and timings were compared with DMRG. Similar accuracies at lower bond
dimensions were obtained with T3NS. For \cuo in a (26e, 44) active space, a comparable accuracy was
obtained at $D=500$ for T3NS and previously published DMRG with DBSS and a maximal bond dimension of
around 2000\cite{Barcza2011}.

For our proof-of-concept, no great effort was made in optimizing the orbital ordering or avoiding
local minima. Furthermore, only the $U(1)$-symmetry of the QC-Hamiltonian was exploited, but no
point group and $SU(2)$-symmetry. One could also consider post-T3NS methods in close similarity to
post-DMRG methods. Some examples are DMRG-SCF\cite{Zgid2008}, DMRG-CASPT2\cite{Kurashige2011} and
DMRG-TCCSD (DMRG-tailored coupled cluster with single and double excitations)\cite{Veis2016}. These
topics will be of interest in future research.

\begin{acknowledgement}
K.G. acknowledges support from the Research Foundation Flanders (FWO Vlaanderen).
\"{O}.L. acknowledges support from the Hungarian National  Research, Development and Innovation
Office (NKFIH) through Grant No. K120569 and the Hungarian Quantum Technology National Excellence
Program (Project No. 2017-1.2.1-NKP-2017-00001).

\end{acknowledgement}

\begin{suppinfo}

The following supporting files are available.
\begin{itemize}
  \item trees.pdf: The used tree tensor networks for the LiF, N$_2$ and \cuo calculations.
\end{itemize}

\end{suppinfo}

\bibliography{jrnlnames,mybib}

\pagebreak

\begin{center}
\textbf{\large Supporting information for: T3NS: three-legged tree tensor network states}
\end{center}

\setcounter{equation}{0}
\setcounter{figure}{0}
\setcounter{table}{0}
\setcounter{page}{1}
\makeatletter
\renewcommand{\theequation}{S\arabic{equation}}
\renewcommand{\thefigure}{S\arabic{figure}}

\begin{figure}[h!] 
  \centering
  \includegraphics[width=\columnwidth]{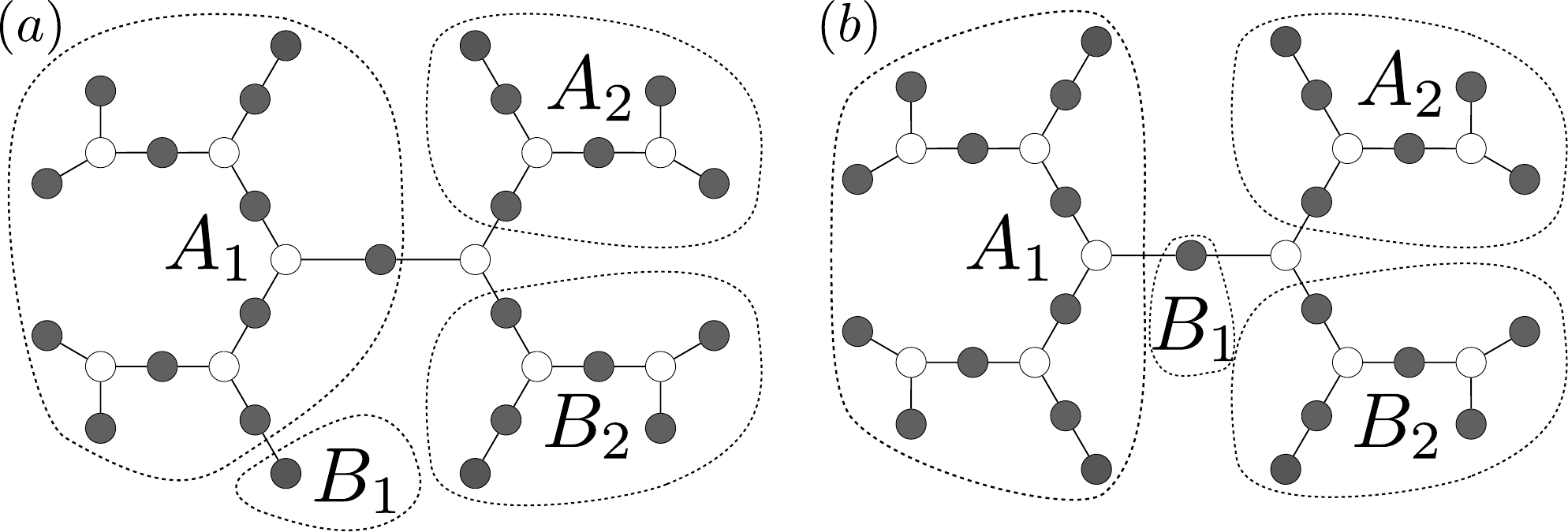}
  \caption{\label{fig:LiForbs} Tree-shaped network for LiF in the (6e, 25) active space. The
    orbitals belonging to the same irreducible representation are grouped. LiF belongs to the
    $C_{2v}$ point group. $A_1, A_2, B_1$ and $B_2$ are the Mulliken symbols of the irreducible
    representations of $C_{2v}$. The orbitals closest to the Fermi level are put as close to the
    center as possible.\\
    (a) is used for the equilibrium bond length $r = 3.05$ a.u. and (b) is used for the two
    calculations at large separation ($r = 12$  a.u. and $r = 13.7$ a.u.).
  }
\end{figure}

\begin{figure}[h!] 
  \centering
  \includegraphics[width=\columnwidth]{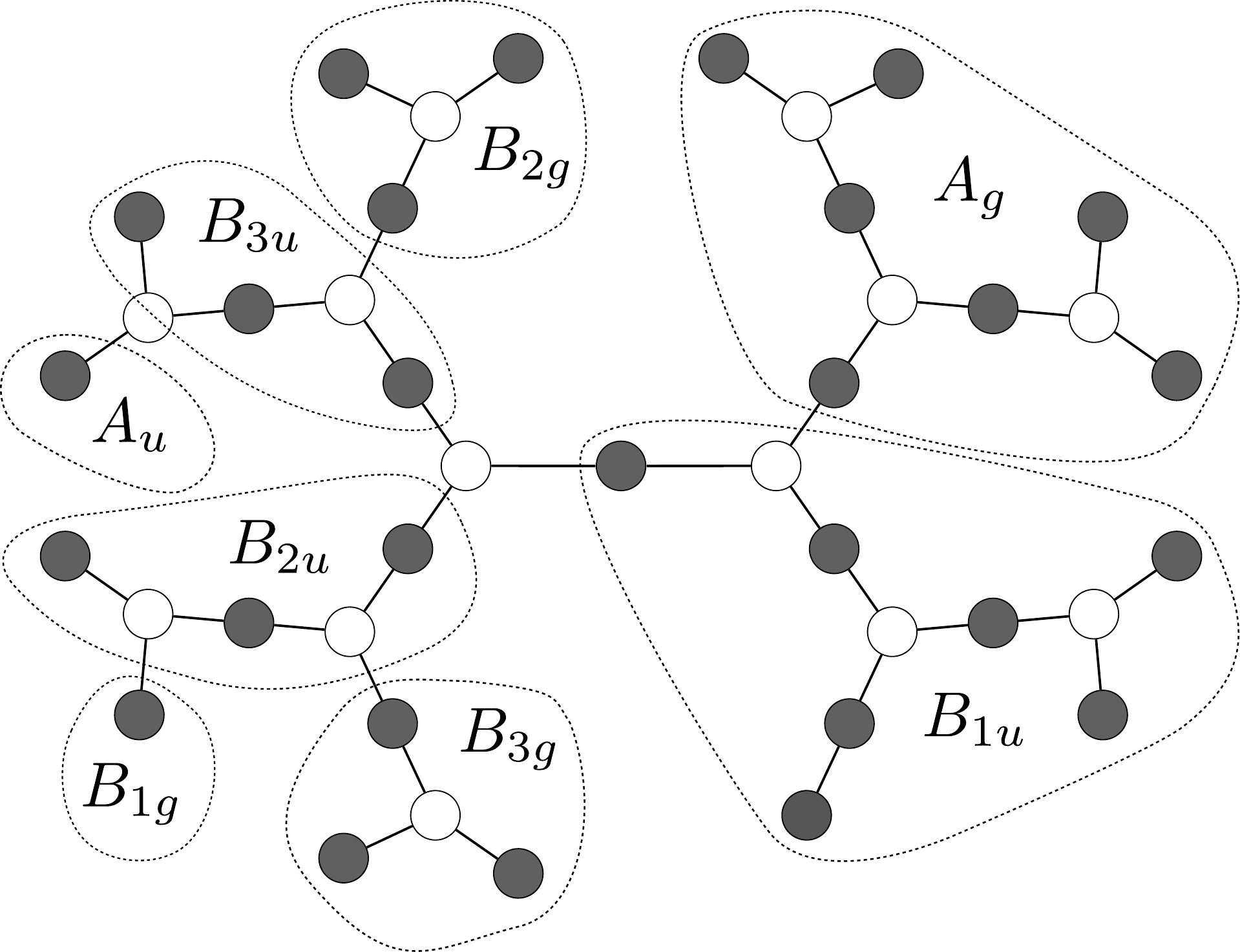}
  \caption{\label{fig:N2orbs} Tree-shaped network for N$_2$ in the cc-pVDZ basis (14e, 28). The
    orbitals belonging to the same irreducible representation are grouped. N$_2$ belongs to the
    $D_{2h}$ point group. $A_g, A_u, B_{1u}, B_{2u}, B_{3u}, B_{1g}, B_{2g}$ and $B_{3g}$ are the
    Mulliken symbols of the irreducible representations of $D_{2h}$. The orbitals closest to the
    Fermi level are put as close to the center as possible. Bonding and anti-bonding irreps are put
    close together.}
\end{figure}

\begin{figure}[h!] 
  \centering
  \includegraphics[width=\columnwidth]{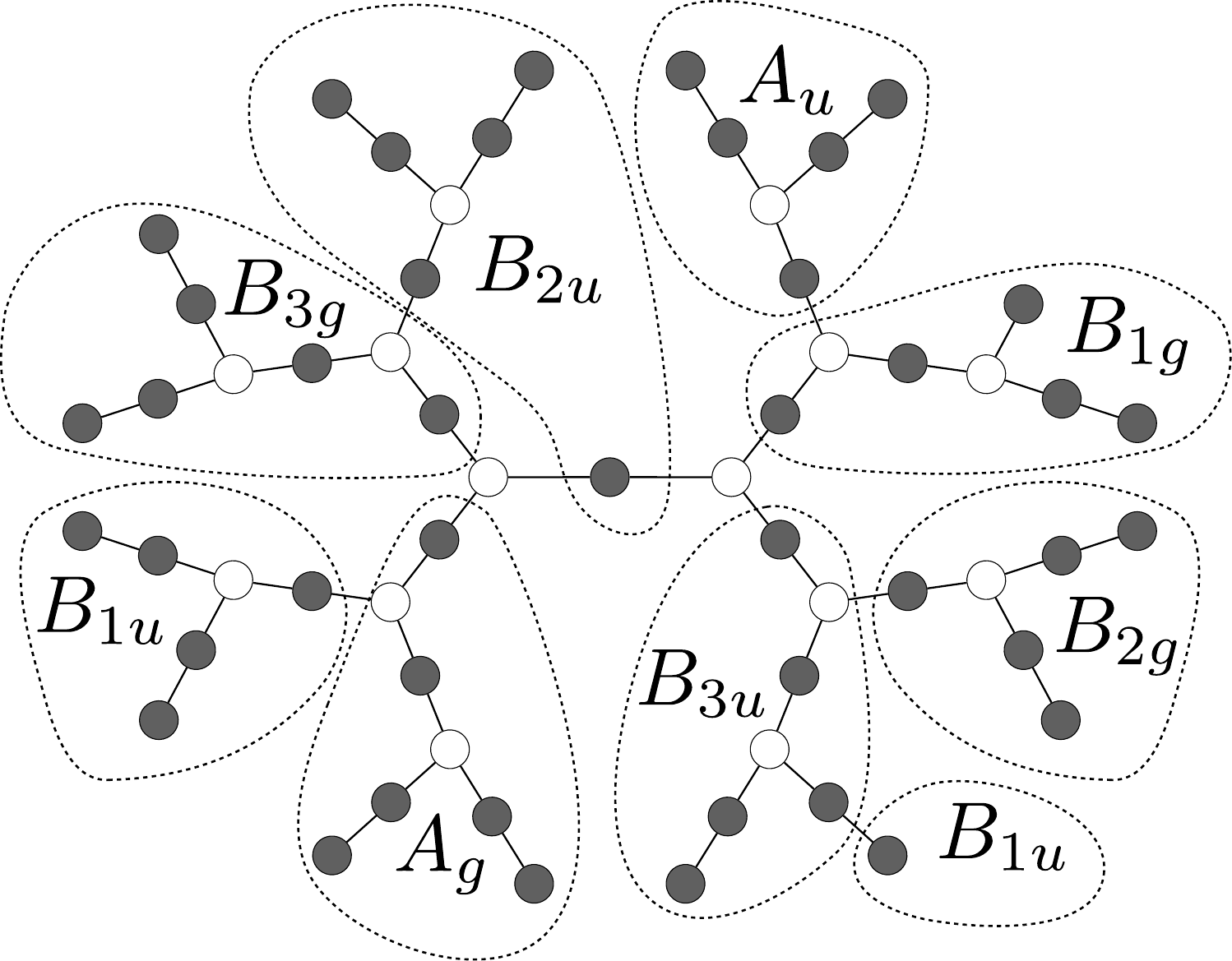}
  \caption{\label{fig:cu2o2orbs} Tree-shaped network for \cuo in the (26e, 44) active space for
    both isomers. The orbitals belonging to the same irreducible representation are grouped. \cuo
    belongs to the $D_{2h}$ point group. $A_g, A_u, B_{1u}, B_{2u}, B_{3u}, B_{1g}, B_{2g}$ and
    $B_{3g}$ are the Mulliken symbols of the irreducible representations of $D_{2h}$. The orbitals
    with the highest single-orbital entropy  are put as close to the center as possible (entropies
    obtained from figs. 2 and 11 of ref.  \citenum{Barcza2011}). Bonding and anti-bonding irreps are
    put close together.}
\end{figure}

\end{document}